\documentclass[aps,prl,floatfix,twocolumn]{revtex4}
\usepackage{graphics}
\usepackage{dcolumn}
\usepackage{epsfig}

\newcommand{\bc}{{\bf c }}

\newcommand{\bk}{{\bf k }}

\newcommand{\bp}{{\bf p }}

\newcommand{\br}{{\bf r }}

\newcommand{\bv}{{\bf v }}

\newcommand{\bG}{{\bf G}}

\newcommand{\bK}{{\bf K}}

\begin{document}

\bibliographystyle{prsty}

\pagenumbering{arabic}

\title
{New Generation of Massless Dirac Fermions\\ in Graphene under External Periodic Potentials}
\author{Cheol-Hwan Park$^{1,2}$}
\email{cheolwhan@civet.berkeley.edu}
\author{Li Yang$^{1,2}$}
\author{Young-Woo Son$^{3,4}$}
\author{Marvin L. Cohen$^{1,2}$}
\author{Steven G. Louie$^{1,2}$}
\affiliation{$^1$Department of Physics, University of California at Berkeley,
Berkeley, California 94720\\
$^2$Materials Sciences Division, Lawrence Berkeley National Laboratory,
Berkeley, California 94720\\
$^3$Department of Physics, Konkuk University, Seoul 143-701, Korea\\
$^4$School of Computational Sciences, Korea Institute for AdVanced Study,
Seoul 130-722, Korea}

\date{\today}

\begin{abstract}
We show that new massless Dirac fermions are
generated when a slowly varying periodic potential is applied to graphene.
These quasiparticles, generated near the supercell Brillouin
zone boundaries with anisotropic group velocity, are different
from the original massless Dirac fermions.
The quasiparticle
wavevector (measured from the new Dirac point),
the generalized pseudospin vector, and the group velocity
are not collinear.
We further show that with an appropriate periodic potential of triangular symmetry,
there exists an energy window over which
the only available states are these quasiparticles, thus,
providing a good system to probe experimentally the new massless Dirac fermions.
The required parameters of external potentials are within the realm of
laboratory conditions.
\end{abstract}

\maketitle

Semiconducting and metallic
superlattice structures are now routinely used in manipulating
the electronic structure of
materials~\cite{tsu:2005SL_to_NEs}.
These superlattices have additional
electronic band gaps at the supercell
Brillouin zone (SBZ) boundary,
which often give rise to
interesting phenomena.

Since the successful isolation of
graphene~\cite{novoselov:2005PNAS_2D,novoselov:2005Nat_Graphene_QHE,
zhang:2005Nat_Graphene_QHE,berger:2006Graphene_epitaxial},
numerous studies have been performed
on this novel material~\cite{geim:2007NatMat_Graphene_Review}.
In particular, there have been a number of interesting theoretical predictions
on graphene superlattices (defined to be graphene under an external
periodic potential or graphene with periodic defects).
For example, for an one-dimensional (1D)
or a two-dimensional (2D) rectangular graphene superlattice,
the group velocity of the low-energy
charge carriers is renormalized
anisotropically~\cite{park:2008NatPhys_GSL};
a corrugated graphene sheet
is expected to show charge
inhomogeneity and localized
states~\cite{guinea:075422};
and arrays of anti-dots (missing carbon atoms) of specific design could
induce band gaps~\cite{pedersen:2008PRL_GSL} or
magnetism~\cite{yu:2008arxiv_GSL}.

Graphene superlattices are not only of theoretical interest,
but have also been experimentally realized. Superlattice
patterns with periodicity as small as 5~nm have been imprinted
on graphene through electron-beam induced deposition of
adatoms~\cite{meyer:123110}.
Also, triangular patterns with $\sim$10~nm
lattice period have been
observed for graphene on metal
surfaces~\cite{marchini:2007PRB_Graphene_Ru,vazquez:2008PRL_Graphene_SL,
pan:2007condmat_Graphene_SL}.
Using periodically patterned gates
is another possible route to make graphene superlattices.

In this paper, we show that when a periodic potential is applied
to graphene, a new generation of massless Dirac fermions is formed
at the SBZ boundaries. The electronic wavevector (measured from the new Dirac point),
the group velocity and a {\it generalized} pseudospin vector, defined below,
of the newly generated massless Dirac fermions
are not collinear anymore.
In 1D or 2D rectangular graphene
superlattices, the features of these new massless Dirac
fermions are obscured by other states existing around the new Dirac point energy.
We show however that, in triangular graphene superlattices (TGSs),
there can be no states other than those of the new massless Dirac
fermions around the energy of the new Dirac points.
Therefore, doped or gated TGSs should provide a clear way to probe this new class of
massless Dirac fermions that are absent in pristine graphene.

A physical requirement for the discussed phenomenon is that
the variation of the external periodic
potential is much slower than the inter-carbon distance
so that inter-valley scattering (between $\bK$ and $\bK'$) may be
neglected~\cite{ando:1998JPSJ_NT_Backscattering,mceuen:1999PRL_NT_Backscattering},
and we limit our discussion to
the low-energy electronic states of graphene which have wavevectors
close to the $\bK$ point.
The Hamiltonian of the low-energy quasiparticles in pristine
graphene in a pseudospin basis,
$\left(\begin{array}{cc}1\\0\end{array}\right)e^{i\bk\cdot\br}$
and $\left(\begin{array}{cc}0\\1\end{array}\right)e^{i\bk\cdot\br}$
(where $\left(\begin{array}{cc}1\\0\end{array}\right)$ and
$\left(\begin{array}{cc}0\\1\end{array}\right)$ are Bloch sums
of $\pi$-orbitals with wavevector $\bK$ on the sublattices $A$ and $B$,
respectively, and $\bk$ is the wavevector from the $\bK$ point),
is given by~\cite{wallace:1947PR_BandGraphite}
\begin{equation}
H_0=\hbar v_0\left(-i\sigma_x\partial_x-i\sigma_y\partial_y\right)
\,,
\label{eq:H_0}
\end{equation}
where $v_0$ is the group velocity
and $\sigma$'s are the Pauli matrices.
The eigenstates and the energy eigenvalues are given by
\begin{equation}
\psi^0_{s,\bk}(\br)=\frac{1}{\sqrt{2}}\left( \begin{array}{c}
1\\
se^{i\theta_\bk}\end{array}\right)e^{i\bk\cdot\br}
\label{eq:solH0}
\end{equation}
and
\begin{equation}
E^0_{s}(\bk)=s\hbar v_0 k\,,
\label{eq:E}
\end{equation}
respectively, where $s=\pm1$ is the band index and $\theta_\bk$ is the
polar angle of the wavevector $\bk$.
Equation~(\ref{eq:solH0}) indicates that the pseudospin vector
is parallel and anti-parallel to the wavevector $\bk$
in the upper band ($s=1$) and in the lower band ($s=-1$), respectively.
Moreover, the pseudospin vector is always parallel to
the group velocity.

Let us first consider the case that a 1D potential $V(x)$, periodic along the $x$
direction with periodicity $L$,
is applied to graphene. The Hamiltonian $H$ then reads
\begin{equation}
H=\hbar v_0\left(
-i\sigma_x\partial_x-i\sigma_y \partial_y+
{I}\ V(x)/\hbar v_0\right),
\label{eq:H}
\end{equation}
where $I$ is the $2\times2$ identity matrix.
After a similarity transform, $H'=U_1^\dagger HU_1$, using the unitary
matrix
\begin{equation}
U_1=\frac{1}{\sqrt{2}}\left(\begin{array}{cc}
e^{-i\alpha(x)/2} & -e^{i\alpha(x)/2}\\
e^{-i\alpha(x)/2} & e^{i\alpha(x)/2}
\end{array}\right)
\label{eq:U_1}
\end{equation}
where $\alpha(x)$ is given by~\cite{note:alpha_v}
\begin{equation}
\alpha(x) = 2\int_0^xV(x')\,dx'/\hbar v_0\,,
\label{eq:alpha}
\end{equation}
we obtain~\cite{note:novikov}
\begin{equation}
H'=\hbar v_0\left(\begin{array}{cc}
-i\partial_x & -e^{i\alpha(x)}\partial_y\\
e^{-i\alpha(x)}\partial_y & i\partial_x
\end{array}\right)\,.
\label{eq:H_prime}
\end{equation}

To obtain the eigenstates and energy eigenspectrum of
$H'$ in general, using a plane wave spinor basis set,
we need an infinite number of plane waves
with wavevectors different from one another by the reciprocal
lattice vectors of the superlattice.
(A reciprocal vector of the superlattice
is given by $\bG_m=m\left({2\pi}/{L}\right)\hat{x}\equiv m\,G_0\, \hat{x}$
where $m$ is an integer.)
However, if we are interested only in quasiparticle states
whose wavevector
$\bk\equiv\bp+\bG_m/2$ is such that
$|\bp|\ll G_0$,
we could treat the terms containing $\partial_y$ in Eq.~(\ref{eq:H_prime})
as a perturbation since $\bG_m$ is along $\hat x$.
$H'$ may be reduced to a
$2\times2$ matrix using the following two states as basis functions
\begin{equation}
\left(\begin{array}{c}
1\\
0
\end{array}\right)'e^{i(\bp+\bG_m/2)\cdot\br}\ {\rm and }\ \
\left(\begin{array}{c}
0\\
1
\end{array}\right)'e^{i(\bp-\bG_m/2)\cdot\br}\,.
\label{eq:basis}
\end{equation}
[Note that the spinors
$\left(\begin{array}{cc}1\\0\end{array}\right)'$ and
$\left(\begin{array}{cc}0\\1\end{array}\right)'$ now have
a different meaning from
$\left(\begin{array}{cc}1\\0\end{array}\right)$ and
$\left(\begin{array}{cc}0\\1\end{array}\right)$
that were defined before because of the unitary transformation.]

In order to calculate these matrix elements, we expand $e^{i\alpha(x)}$ as
\begin{equation}
e^{i\alpha(x)}=
\sum^{\infty}_{l=-\infty}f_l[V]e^{\,i\,l\,G_0\,x},
\label{eq:e_alpha}
\end{equation}
where $f_l[V]$'s are coefficients determined by the
periodic potential $V(x)$.
One important thing to note is that in general
\begin{equation}
|f_l|<1\,,
\label{eq:f}
\end{equation}
which can directly be deduced from Eq.~(\ref{eq:e_alpha}).
The physics simplifies when the external potential $V(x)$ is an even
function.
Then, $f_l[V]$'s are all real~\cite{note:even}.
For states with wavevector $\bk$ very close to $\bG_m/2$, the $2\times2$ matrix $M$
whose elements are calculated from the Hamiltonian $H'$ with
the basis given by Eq.~(\ref{eq:basis})
can be written as
\begin{equation}
M=\hbar v_0\left(p_x\sigma_z+f_m p_y\sigma_y\right)
+\hbar v_0\,mG_0/2\cdot I
\,.
\label{eq:M}
\end{equation}
After performing yet another similarity transform
$M'=U_2^\dagger MU_2$
with
\begin{equation}
U_2=\frac{1}{\sqrt{2}}\left(\begin{array}{cc}
1 & 1\\
-1 & 1
\end{array}\right)\,,
\label{eq:U_2}
\end{equation}
we obtain the final result:
\begin{equation}
M'=\hbar v_0\left(p_x\sigma_x+f_m\, p_y\sigma_y\right)
+\hbar v_0\,m\,G_0/2\cdot I
\,.
\label{eq:M_prime}
\end{equation}
The only difference of the Hamiltonian in Eq.~(\ref{eq:M_prime})
from that in Eq.~(\ref{eq:H_0}),
other than a constant energy term,
is that the group velocity
of quasiparticles moving along
the $y$ direction has been changed from $v_0$ to
$f_m v_0$~\cite{note:genWeyl}.
Thus, the electronic states near $\bk=\bG_m/2$ are also those of
massless Dirac fermions but having a group velocity
varying {\it anisotropically} depending on the propagation direction.
Moreover, the group velocity along the $y$ direction
is {\it always lower} than $v_0$ [Eq.~(\ref{eq:f})] regardless of the form or magnitude of the
periodic potential~$V(x)$.

  \begin{figure}
  \includegraphics[width=0.6\columnwidth]{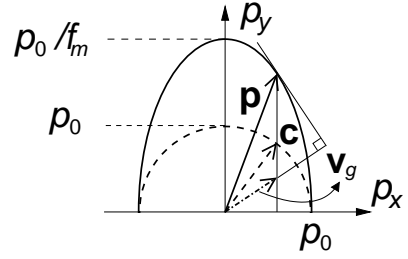}
  \caption{
  Schematic diagram showing an equi-energy contour (ellipse)
  with $E=\hbar v_0 k_0+\hbar v_0\,m\,G_0/2$
  of the newly generated massless Dirac fermions.
  The quasiparticle wavevector $\bk$, the generalized
  pseudospin vector (see text) $\bc$,
  and the group velocity vector $\bv_g$ are represented by
  solid, dashed and dash-dotted arrows, respectively, for graphene
  in an even periodic potential.}
  \label{Fig1}
  \end{figure}

The eigenstate and the energy eigenvalue of the matrix $M'$ are given by
\begin{equation}
\varphi_{s,\bp}=\frac{1}{\sqrt{2}}\left( \begin{array}{c}
1\\
se^{i\phi_\bp}\end{array}\right)''
\label{eq:phi}
\end{equation}
and
\begin{equation}
E_{s}(\bp)=s\hbar v_0 \sqrt{p_x^2+|f_m|^2p_y^2}+\hbar v_0\, m\, G_0/2\,,
\label{eq:E_p}
\end{equation}
respectively, where $\phi_\bp$ is the
polar angle of the pseudospin vector {\bf c} of $\varphi_{s,\bp}$,
which is parallel to $s(p_x\hat{x}+f_m\, p_y\hat{y})$.
The spinor $\varphi_{s,\bp}$, however, should not be confused with
the one in Eq.~(\ref{eq:solH0}) representing the sublattice degree of freedom,
or with the one in Eq.~(\ref{eq:basis}). A double prime
in Eq.~(\ref{eq:phi}) emphasizes this point.

The eigenstate $\psi_{s,\bk}(\br)$ of the original Hamiltonian $H$
in Eq.~(\ref{eq:H}) can be obtained
by using Eqs.~(\ref{eq:U_1}), (\ref{eq:basis}), (\ref{eq:U_2}) and (\ref{eq:phi}).
Since the unitary
transforms conserve the inner-product between eigenstates,
if a {\it generalized} pseudospin vector for the original Hamiltonian $H$
in Eq.~(\ref{eq:H}) is defined as the pseudospin vector of
the transformed Hamiltonian $M'$, i.\,e.\,, {\bf c},
the scattering matrix elements between states of these new massless Dirac fermions
due to long-wavelength perturbations are
described by the generalized pseudospin
in the same manner as those of the original
massless Dirac fermions in pristine graphene are described
with their pseudospin.

On the other hand, the group velocity vector $\bv_g$ is parallel to
$s\,(p_x\hat{x}+f_m^2\, p_y\hat{y})$ [Eq.~(\ref{eq:E_p})]. Therefore, in general,
the three vectors $\bp$, $\bc$ and $\bv_g$ are not collinear
(Fig.~\ref{Fig1}).
However, it is obvious that if the wavevectors ($\bp$) of two electronic
states are aligned or anti-aligned to each other, so are their
generalized pseudospin vectors, as in pristine graphene, resulting in a maximum or
a zero overlap between the two states, respectively.
If $V(x)$ is not an even function, the dispersion relation of the new massless Dirac
fermions remains the same as Eq.~(\ref{eq:E_p}) but a generalized pseudospin vector
may not be defined~\cite{note:fgeneral}.

Similarly, for graphene in slowly varying 2D periodic potential,
new massless Dirac fermions are generated centered
around the wavevectors $\bk_c=\bG/2$ where the $\bG$'s are
the superlattice reciprocal vectors.
A state with wavevector $\bk$ around $\bk_c$ mixes strongly
with another state with wavevector $\bk-\bG$ by the superlattice potential.
Applications of the same argument that we made use of in
the case of 1D graphene superlattices result in linear band dispersions.

Even though new massless Dirac fermions are generated
in 1D graphene superlattices, because there is no SBZ
boundary perpendicular to the periodic direction,
they are obscured by other states,
and there is no new value of energy at which the density of states vanishes.
In a 2D rectangular graphene superlattices,
the SBZ is a rectangle. It turns out that
the energy separation at the SBZ corners also vanishes
due to the chiral nature of graphene~\cite{park:2008NatPhys_GSL}.
Therefore, in 2D rectangular graphene superlattices,
again, there are states other than the new massless Dirac fermions
in the range of the new Dirac point energy. However, as we show below, in TGSs,
there can exist an energy window within which the only available states are the newly
generated massless Dirac fermions.

  \begin{figure}
  \includegraphics[width=0.9\columnwidth]{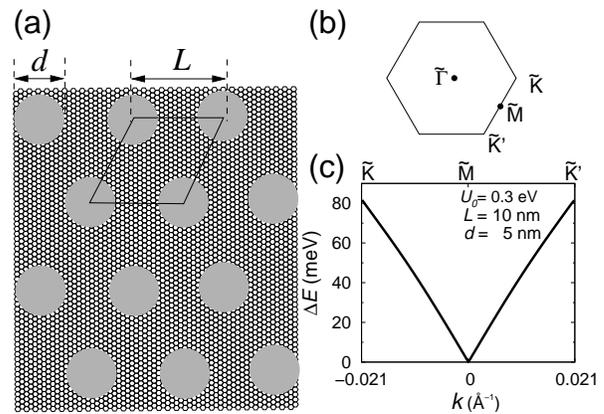}
  \caption{
  (a): A TGS with muffin-tin type of
  periodic potential with a spatial period $L$. The potential is
  $U_0$ inside the gray disks with diameter $d$ and zero outside.
  (b) The SBZ of a TGS.
  (c) The energy separation $\Delta E$ between states in the first and the second band above the
  original Dirac point energy versus the wavevector $k$ along the path
  $\tilde{{\text K}}\tilde{{\text M}}\tilde{{\text K'}}$
  in a TGS given by $U_0=$0.5~eV, $L=10$~nm, and $d=5$~nm.}
  \label{Fig2}
  \end{figure}

As an illustration, we consider a TGS shown in Fig.~\ref{Fig2}(a).
The external potential is of a muffin-tin type
with value $U_0$ in a triangular array of disks of diameter $d$
and zero outside of the disks. The spatial period of the superlattice is $L$.
Figure~\ref{Fig2}(b) shows the SBZ of a TGS.

Figure~\ref{Fig2}(c) shows the electron energy separation between
states in the first and
the second band above the original Dirac point energy along the path
$\tilde{{\text K}}\tilde{{\text M}}\tilde{{\text K'}}$ in the SBZ
[Fig.~\ref{Fig2}(b)] for a TGS.
The energy separation at the corner,
or the $\tilde{\text K}$ point, of the SBZ is largest,
contrary to that of the rectangular graphene superlattices
where the energy separation closes at the SBZ corners~\cite{park:2008NatPhys_GSL};
but that at the $\tilde{\text M}$ point
is zero. New massless Dirac fermions are thus formed
around the $\tilde{\text M}$ points.
With the set of potential parameters in Fig.~\ref{Fig2}
($U_0=$0.5~eV, $L=10$~nm, and $d=5$~nm), the energy separation
at the $\tilde{\text K}$ point is 82~meV,
much larger than room-temperature thermal energy.
This energy separation can be tuned by changing the superlattice parameters.

  \begin{figure}
  \includegraphics[width=0.6\columnwidth]{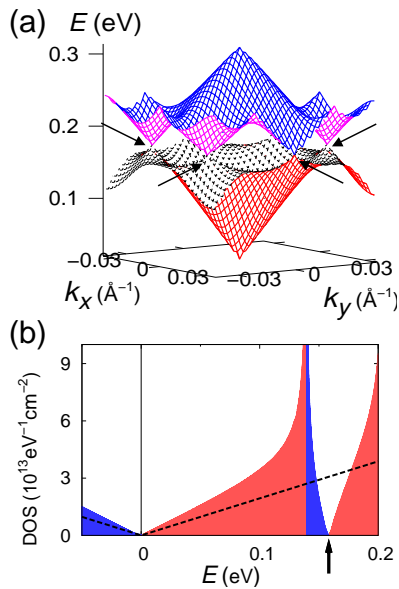}
  \caption{(color online)
  (a): Energy dispersion relation of a TGS with external potential
  with $U_0=0.5$~eV, $L=10$~nm and $d=5$~nm for the first
  and the second band above the original
  Dirac point energy as a function of wavevector $\bk$ from the original Dirac point.
  Arrows indicate the $\tilde{{\text M}}$ points of the SBZ around which
  new massless Dirac fermions are generated.
  (b): The DOS of charge carriers
  in electron orbits (bright and red) and hole orbits (dark and blue)
  in the TGS characterized in (a). The original Dirac point energy is set at zero.
  Dashed black line shows the DOS of pristine graphene.
  The arrow indicates the new Dirac point energy.}
  \label{Fig3}
  \end{figure}

Figure~\ref{Fig3}(a) shows the energy dispersions of the first and the second
band of the considered TGS. We can see the linear
energy dispersion relation at
the $\tilde{{\text M}}$ points [Fig.~\ref{Fig2}(c)].
Close to the original Dirac point energy ($E=0$), the density of states (DOS) varies linearly with
energy, similar to
that of pristine graphene, except that the slope
is larger because of the reduced band velocity.
At around $E=0.16$~eV, there exists another energy value
where the DOS vanishes also linearly.

In conclusion, we have shown that a new class of massless Dirac fermions
are generated in graphene when a periodic potential is applied
and we have studied the novel characteristics of these quasiparticles.
Moreover, in triangular graphene superlattices, there can exist energy windows
where there are no other states than these new quasiparticles.
The triangular graphene superlattices thus should provide a good platform for experimental
probing of the new massless Dirac fermions predicted here.

C.-H.P. thanks Dmitry Novikov for fruitful discussions.
This work was supported by NSF Grant
No. DMR07-05941 and by the Director, Office of Science, Office of Basic Energy
Sciences, Division of Materials Sciences and Engineering Division,
U.S. Department of Energy under Contract No. DE- AC02-05CH11231.
Y.-W.S. was supported by KOSEF grant R01-2007-000-10654-0 and by
Nano R\&D program 2008-03670 through the KOSEF funded by the Korean
government (MEST).
Computational resources have been provided by NPACI and NERSC.

{\it Note added in proof. -} After submission of this Letter,
an angle-resolved photoemission experiment on graphene on Ir(111) surface resulting
in superlattice formation was reported~\cite{pletikosic:Ir} in which minigap openings
at the SBZ boundary are found and evidence of replicas of the primary Dirac cone observed.


\begin{thebibliography}{10}
\bibitem{note:alpha_v}
We assume that appropriate constants are subtracted from $V(x)$ and $\alpha(x)$
  so that the averages of $V(x)$ and $\alpha(x)$ are both zero.
\bibitem{note:novikov}
A similar transform has been performed on the Hamiltonian of a carbon nanotube
  under a sinusoidal potential (Refs.~\onlinecite{PhysRevLett.87.276802} and
  \onlinecite{novikov:235428}). In carbon nanotubes, a finite scattering
  probability along the periodic direction is given by a boundary condition
  along the circumferential direction, curvature or magnetic field
  (Refs.~\onlinecite{PhysRevLett.87.276802} and \onlinecite{novikov:235428}),
  whereas in graphene, the wavevector component $k_y$ naturally serves the role
  of scattering along the periodic direction.
\bibitem{note:even}
If $V(x)$ is even, $\alpha(x)$ is odd [Eq.~(\ref{eq:alpha}) and
  Ref.~\onlinecite{note:alpha_v}]. If we take the complex conjugate of
  Eq.~(\ref{eq:alpha}) and change $x$ to $-x$, it is evident that $f_l[V]$'s
  are real.
\bibitem{note:fgeneral}
If we remove the assumption that the periodic potential is an even function,
  Eqs.~(\ref{eq:M}) and~(\ref{eq:M_prime}) reads $M=\hbar
  v_0\left(\begin{array}{cc} p_x & -if_m p_y\\ if_m^*p_y & -p_x
  \end{array}\right) +\hbar v_0\,mG_0/2\cdot I$ and $ M'=\hbar
  v_0\left(\begin{array}{cc} -p_y\,{\rm Im}{f_m} & p_x-i\,p_y\,{\rm Re}{f_m}\\
  p_x+i\,p_y\,{\rm Re}{f_m} & p_y\,{\rm Im}{f_m} \end{array}\right) +\hbar
  v_0\,m\,G_0/2\cdot I\,, $ respectively. The energy eigenvalue of $M'$ is
  given by Eq.~(\ref{eq:E_p}). The eigenstate of $M'$ corresponding to
  Eq.~(\ref{eq:phi}) is given by $
  \varphi_{s,\bp}=\frac{1}{\sqrt{1+\lambda_{s,\bp}^2}}\left( \begin{array}{c}
  1\\ s\,\lambda_{s,\bp}\,e^{i\phi_\bp}\end{array}\right)''\,, $ where
  $\phi_\bp$ is the polar angle of the vector $s(p_x\hat{x}+p_y\,{\rm
  Re}f_m\,\hat{y})$, and $ \lambda_{s,\bp}=\left[
  \sqrt{p_x^2+|f_m|^2p_y^2}+s\,p_y\,{\rm Im}f_m \right]/
  \sqrt{p_x^2+p_y^2\,({\rm Re}f_m)^2}\,. $ In general, $\lambda_{s,\bp}$ is not
  1, and moreover, varies with the direction of $\bp$.
\end{thebibliography}
\end{document}